\documentclass[12pt]{article}
\usepackage{amsfonts,amsmath,amssymb,epsf}
\usepackage{amsmath}
\usepackage{graphicx}
\usepackage{physics}
\usepackage{comment}
\usepackage{color}
\usepackage{tikz}
\usepackage{graphicx,hyperref,color}
\hypersetup{
    unicode=false,          % non-Latin characters in Acrobat’s bookmarks
    pdftoolbar=true,        % show Acrobat’s toolbar?
    pdfmenubar=true,        % show Acrobat’s menu?
    linktocpage=true,       % hyperlink by page number in contents
    pdffitwindow=false,     % window fit to page when opened
    pdfstartview={FitH},    % fits the width of the page to the window
    pdfnewwindow=true,      % links in new PDF window
    colorlinks=true,       % false: boxed links; true: colored links
    linkcolor=blue,          % color of internal links (change box color with linkbordercolor)
    citecolor=blue,        % color of links to bibliography
    filecolor=blue,      % color of file links
    urlcolor=blue           % color of external links
}

\numberwithin{equation}{section}
\begin{document}

\begin{titlepage}

\thispagestyle{empty}
\vspace*{-2cm}
\begin{flushright}
YITP-24-40 \\
RIMS-1982
\\
~
\\
\end{flushright}
\bigskip
\begin{center}
\noindent{{\Large \textbf{Celestial CFT from $H_3^+$-WZW Model}}}\\
\vspace{2cm}
Naoki Ogawa$^{a}$\footnote{\href{mailto:naoki.ogawa@yukawa.kyoto-u.ac.jp}{\texttt{naoki.ogawa@yukawa.kyoto-u.ac.jp}}},\ Shunta Takahashi$^b$\footnote{\href{mailto:shunta@kurims.kyoto-u.ac.jp}{\texttt{shunta@kurims.kyoto-u.ac.jp}}},\ Takashi Tsuda$^a$\footnote{\href{mailto:takashi.tsuda@yukawa.kyoto-u.ac.jp}{\texttt{takashi.tsuda@yukawa.kyoto-u.ac.jp}}}, and Takahiro Waki$^a$\footnote{\href{mailto:takahiro.waki@yukawa.kyoto-u.ac.jp}{\texttt{takahiro.waki@yukawa.kyoto-u.ac.jp}}}
\setcounter{footnote}{0}
\vspace{1cm}\\
{\it $^a$Center for Gravitational Physics and Quantum Information,\\
Yukawa Institute for Theoretical Physics, Kyoto University, \\
Kitashirakawa Oiwakecho, Sakyo-ku, Kyoto 606-8502, Japan}\\
\vspace{1mm}
{\it $^b$Research Institute for Mathematical Science, Kyoto University}\\
\bigskip \bigskip
\vskip 2em
\end{center}

\begin{abstract}
    Recently, there has been a growing interest in celestial holography, which is holography in asymptotically flat spacetimes. This holographic duality exhibits numerous mysterious and fruitful features, particularly on the dual CFT side.
    In this paper, we present the candidate of dual CFT for Minkowski spacetime extracted from $SL\qty(2,\mathbb{C})/SU\qty(2)\cong H^+_3$ Wess-Zumino-Witten (WZW) model, the simplest non-compact CFT. We demonstrate that it reproduces the well-known principal series and correlation functions dual to the bulk scattering amplitudes.

\end{abstract}

\end{titlepage}

\newpage

\tableofcontents

\clearpage
%%%%%%%%%%%%%%%%%%%%%%%%%%%%%%%%
\section{Introduction}
In recent years, an important extension of the holographic principle \cite{tHooft:1993dmi, Susskind:1994vu} known as celestial holography has been proposed, which is a correspondence in asymptotically flat spacetimes \cite{Pasterski:2016qvg, Cardona:2017keg, Strominger:2017zoo, Pasterski:2017kqt, Pasterski:2017ylz, Donnay:2020guq}. This proposal asserts that quantum gravity in four-dimensional asymptotically flat spacetimes is equivalent to conformal field theory (CFT) on a two-dimensional sphere. The central object of this correspondence is the relationship between scattering amplitudes in four-dimensional spacetime and correlation functions in the two-dimensional CFT (Celestial CFT; CCFT). While there are some suggestions about the non-unitary nature of CCFT \cite{Pasterski:2022lsl, Ogawa:2022fhy}, many details remain unclear.

In the context of holographic principle, the location where CFT is defined is an important element. In AdS/CFT correspondence, CFT is defined on the asymptotic boundary of the AdS spacetime, and it is understood that the position of this boundary, the IR cutoff of the spacetime, corresponds to the UV cutoff in CFT. However, since celestial holography is codimension-two holography, the location where CFT is defined becomes non-trivial. Therefore, this leads to various proposals
such as foliating Minkowski space into (A)dS slices and applying a holographic dictionary to each slice \cite{deBoer:2003vf, Cheung:2016iub} . We infer the existence of CFT on the asymptotic boundary of the on-shell hyperboloid (Euclidean AdS${}_3$, $H^+_3$) in momentum space from the fact that correlators can be expressed as Witten diagrams on it \cite{Pasterski:2016qvg,Cardona:2017keg,Casali:2022fro,Iacobacci:2022yjo}.

Other than celestial holography, we have a very concrete example of holographic principle in AdS/CFT.
Recently, a specific setup of AdS${}_3$/CFT${}_2$, namely AdS${}_3$ string / symmetric orbifold duality, has been proposed in some specific cases, 
Eberhardt dual (unit NSNS flux) \cite{Gaberdiel:2018rqv,Eberhardt:2018ouy,Eberhardt:2019ywk,Eberhardt:2021vsx},  extension to more NSNS flux case \cite{Eberhardt:2019qcl,Hikida:2023jyc}, less than unit flux case \cite{Balthazar:2021xeh}, for example.
In those cases, correlation functions from both sides coincide.
AdS${}_3$ side is a string theory on Lorentzian AdS${}_3$ with some background $B$-field, and its principal part of the world-sheet theory is $SL(2,\mathbb{R})_k$ Wess-Zumino-Witten (WZW) model. Its Euclidean AdS${}_3$ counterpart $SL\qty(2,\mathbb{C})/SU\qty(2)$ WZW model, also known as $H_3^+$-WZW model, are investigated in detail by Teschner \cite{Teschner:1997ft, Teschner:1997fv, Teschner:1999ug}.
On the other hand, CFT${}_2$ side is so-called symmetric orbifold CFT. 
In this duality, the $J_0$ eigenvalue in % $H_3^+$-
WZW model is considered as conformal weight in CFT side.
We will review this kind of AdS${}_3$/CFT${}_2$ only briefly, but
this is also one of the important concepts of this paper.

In this paper, we propose $H_3^+$-WZW model as a toy-model of CCFT in the following logic.
Firstly, we assume the existence of dual boundary CFT of $H_3^+$-WZW string theory as done in AdS${}_3$ string / symmetric orbifold duality. 
Correlation functions of this dual boundary CFT can be evaluated from $H^+_3$-WZW model, as explicitly done in section \ref{sec: CalcCorrelation} for the case of two-point functions and three-point functions.
These two-/three- point correlators are shown to have exactly the expected form in CCFT,
and thus we claim that the dual boundary CFT of $H_3^+$-WZW string theory can be interpreted as CCFT. In our setup, the corresponding theory in four-dimensional Minkowski spacetime is massive scalar theory.

When considering
$H_3^+$-WZW string theory in so-called mini-superspace limit, bulk excitation becomes geodesic and one can expect that our CCFT construction can be understood in semi-classical AdS${}_3$/CFT${}_2$ point of view.
More concretely, operators of our interest correspond to excitations that violate BF bound in traditional AdS${}_3$/CFT${}_2$ setup, as explained in section \ref{sec: 4.5}.
Actually, we can start from semi-classical AdS${}_3$/CFT${}_2$ and construct CCFT toy-model without considering any stringy setup.
% $\Delta = 1 + \sqrt{1+(m/R_{\text{AdS}})^2}$
Although 
this procedure is one of our important results and supports our main claim, it could not be incorporated into the logical flow of the main argument of this paper, thus we decided to include it in appendix \ref{classical}.

This paper is organized as follows. 
In section \ref{sec: ccft_review}, we first provide a brief review of celestial holography. Subsequently, we consider celestial holography as AdS/CFT in momentum space. Specifically, we introduce that correlation functions in CCFT can be expressed in the form of Witten diagrams on the momentum hyperboloid.
In section \ref{sec: WZW_Review}, we provide a short review of $H^+_3$-WZW model and AdS$_3$ string / symmetric orbifold CFT.
In section \ref{sec: CalcCorrelation}, we introduce the dual boundary CFT of $H^+_3$-WZW string theory and propose it as a toy-model of CCFT. We compare both spectrum, two-point functions, and three-point functions to examine validity as the toy-model of CCFT of massive scalar theory.
In section \ref{sec : CandD}, we conclude this paper and give some future directions.
We present an alternative (but understandable as a limit of our model) CCFT toy-model construction from semi-classical AdS${}_3$/CFT${}_2$ in appendix \ref{classical}.

%%%%%%%%%%%%%%%%%%%%%%%%%%%%%%%%
\section{Review of celestial CFT} \label{sec: ccft_review}
In this section, we review the fundamental aspects of celestial holography.
%we give some comment about where CCFT is defined in the context of Witten diagram.
In section \ref{CCFT_momentum}, we review that the CCFT correlation function can be written as a Witten diagram on EAdS${}_3$ (momentum hyperboloid). This fact suggests that CCFT is holographic dual of EAdS${}_3$.
This section is partially based on the nice reviews for celesital holography \cite{Raclariu:2021zjz, Pasterski:2021rjz}.

\subsection{Conformal primary operators and correlation functions in CCFT}
\label{section 2.1}
\begin{figure}[htbp]
            \centering
    
            \begin{tikzpicture}[scale = 0.8]
              \draw[->,>=stealth,semithick] (-4,0)--(4,0) node[right]{$\vec{\hat{p}}$}; % x-axis
              \draw[->,>=stealth,semithick] (0,-1)--(0,5) node[left]{$p^0$}; % t-axis
              \draw (0, 0) node[below left]{O}; % the origin
              \draw (0, 1) node[below left]{$1$};
              \draw[very thick, samples = 100, domain = -2.1:2.1] plot({(exp(\x)-exp(-\x))/2}, {(exp(\x)+exp(-\x))/2});
              \draw[dashed, samples = 100, domain = -1:4] plot(\x, \x);
              \draw[dashed, samples = 100, domain = -4:1] plot(\x, -\x);
              \coordinate[label = above left:$e^{i\hat{p}\cdot X}$](A)at({(exp(1)-exp(-1))/2}, {(exp(1)+exp(-1))/2});
              \fill[black](A)circle(0.07);
            \end{tikzpicture}
    
            \caption{unit mass hyperboloid}
    \label{unit mass hyperboloid}
\end{figure}
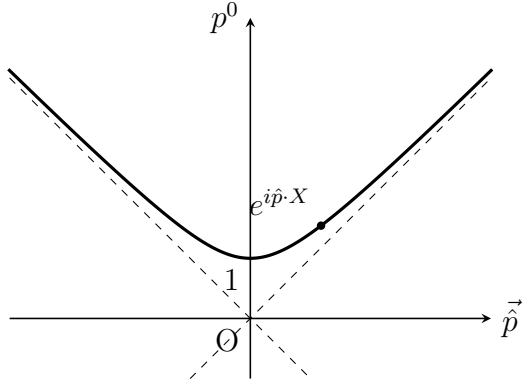

In celestial holography, scattering amplitudes in four-dimensional asymptotic flat spacetime corresponds to correlation functions in CCFT, \textit{Celestial Amplitudes}. 
This correspondence can be understood as a transformation of the basis in scattering amplitudes, which is termed as the \textit{conformal basis}. This conformal basis is built from \textit{conformal primary wave functions} $\Psi_{\Delta}(x ; z)$, which can be expressed using plane waves as follows \footnote{
Here, the measure of hyperbolic integral is defined as follows:
\begin{align*}
    \int_{H_3} [d^3 \hat{p}]=\int \frac{dy}{y^3}\int d^2z. \label{conformal_primary}
\end{align*}
}\cite{Raclariu:2021zjz}:
\begin{align}
    \Psi_{\Delta}(x ; w)=\int_{H_3} [d^3 \hat{p}]~ G_{\Delta}(\hat{p} ; w) e^{i m \hat{p} \cdot x},
\end{align}
where $m$ is mass of massive scalar field.
Here, $G_\Delta(\hat{p};z)$ is a bulk-boundary propagator of the Euclidean AdS$_3$/CFT$_2$ correspondence:
\begin{align} \label{bulk_boundary}
    G_\Delta(\hat{p}(y,z);w)=\frac{1}{(-\hat{p}(y,z)\cdot q(w))^{\Delta}}=\left( \frac{y}{y^2+|z-w|^2} \right)^{\Delta},
\end{align}
where $(y,z): y>0, z\in\mathbb{C}$ is the parameter of on-shell momentum:
\begin{align}
    \hat{p}^\mu(y,z)=\frac{1}{2y}(1+y^2+z\bar{z}, z+\bar{z},i(\bar{z}-z),1-y^2-z\bar{z}),~~~~\hat{p}^2=-1, \label{on-shell momentum}
\end{align}
and $q(w): w\in\mathbb{C}$ is null vector:
\begin{align}
    q^\mu(w)=(1+w \bar{w}, w+\bar{w}, i(\bar{w}-w), 1-w \bar{w}),~~~~q^2=0 .
\end{align}
We should note that this $(y,z)$ corresponds to the point on the unit mass hyperboloid in momentum space (see Fig.\ref{unit mass hyperboloid}). 

Using this conformal primary wave function, we can define the operator of CCFT
\begin{align}
    \mathcal{O}_{\Delta}(z)=(\Psi_{\Delta}(x ; z),\phi(x))_{\text{KG}},
\end{align}
where $\phi(x)$ is a field operator in four-dimensional spacetime, and $(,)_{\text{KG}}$ is a Klein-Gordon inner product.
For example, correlation functions correspond to the scattering amplitude $\mathcal{A}$ for $k \to l$ particles of massive scalar fields is given by \cite{Furugori:2023hgv}:
\begin{align}
   &\left\langle \prod_{i=1}^l \mathcal{O}^-_{1+i\lambda_i} (z_i)\prod_{j=1}^k \mathcal{O}^+_{1+i\lambda_j}(z_j)\right\rangle \nonumber\\
   &= \prod_{i=1}^l \prod_{j=1}^k\int_{H_3} [d\hat{p}^\text{out}_i]c_iG_{1+i\lambda_i}(\hat{p}_i^{\text{out}}; z_i)\int_{H_3} [d\hat{p}^\text{in}_j] c_j^{\ast}G_{1+i\lambda_j}(\hat{p}_j^{\text{in}}; z_j)\,i\mathcal{A}(p^\text{out}_1,...,p^\text{out}_l; p^\text{in}_1,...,p^\text{in}_k),
   \label{massive-dict-amp}
\end{align}
where $c_j:=2^{-i\lambda_j}m_j^{1+i\lambda_j} \sqrt{\lambda_j}/(2\pi)^\frac{5}{2}$ and $\lambda\in\mathbb{R}_{>0}$. This restriction for the conformal dimension $\Delta\in1+i\mathbb{R}_{>0}$ is called \textit{principal series}, which is derived from normalizable condition for conformal primary wavefunctions \cite{Pasterski:2017kqt}. For massless fields, this is given as $\Delta\in1+i\mathbb{R}$.
\footnote{Note that this relation is modified dictionary introduced in \cite{Furugori:2023hgv}. In massive case, this is same as ordinary dictionary (e.g.\cite{Pasterski:2021rjz, Raclariu:2021zjz}) up to normalization factor. }
We introduce shadow transformation as follows:
\begin{align}
    \widetilde{\mathcal{O}_\Delta}(z):=\frac{\Delta-1}{2\pi}\int d^2 z^{\prime}\frac{1}{|z-z^{\prime}|^{2(2-\Delta)}}\mathcal{O}_{\Delta}(z').
    \label{shadow}
\end{align}
Though there are some choice of the notations for the normalization factor, we take the above notation in order to return to the original form after taking this transformation twice:
\begin{align}
    \widetilde{\widetilde{\mathcal{O}_\Delta}}(z)=\mathcal{O}_{\Delta}(z).
\end{align}
Although there is still ambiguity in phases, this will be fixed later.
In the above notation, we can compute two-point functions from passing though amplitudes \cite{Furugori:2023hgv}:
\begin{align}
 &\ev*{\mathcal{O}^-_{1+i\lambda_1} (z_1)\mathcal{O}^+_{1+i\lambda_2}(z_2)}=
    \frac{ \delta(\lambda_1-\lambda_2)}{|z_1-z_2|^{2(1+i\lambda_1)}}, \label{nonshadow_nonshadow}
    \\
     & \ev*{\mathcal{O}^-_{1+i\lambda_1} (z_1)\widetilde{\mathcal{O}^+_{1+i\lambda_2}}(z_2)}=\frac{2\pi i}{\lambda_1}\delta(\lambda_1-\lambda_2)\delta^{(2)}(z_1-z_2). \label{nonshadow_shadow}
\end{align}
From momentum conservation for scattering amplitude, massive three-point CCFT correlators corresponding to scattering amplitude for $m\to m+m$ should vanish:
\footnote{
Massless three-point correlators are known to have a distributional behavior. It has a support only in collinear limit. See e.g. \cite{Law:2019glh}. However, massive three-point functions ($m\rightarrow m+m$) vanish in any limit. Therefore, massive three-point function is exactly $0$ for any momenta.
}
\begin{align}
    \ev{\mathcal{O}_{\Delta_1}\mathcal{O}_{\Delta_2}\mathcal{O}_{\Delta_3}}=0.
\end{align}

\subsection{Celestial CFT on momentum hyperboloid}
\label{CCFT_momentum}
In this subsection, we would like to mention the reason why we state that CCFT may be holographic dual CFT of momentum space hyperboloid (EAdS${_3}$).
It is known that celestial amplitudes can be expressed as correlators on EAdS${_3}$ \cite{Pasterski:2016qvg, Cardona:2017keg}. For concreteness, we consider massive three-point amplitudes with interaction $L_{int}=g\phi\psi^2$ where $\phi$ is massive scalar field with mass $2m(1+\varepsilon)$, and $\psi$ is massive scalar field with mass $m$. In this case, celestial amplitude $\tilde{\mathcal{A}}_3$ which corresponding to the tree-level three-point scattering amplitude $\phi\rightarrow\psi\psi$ is calculated in \cite{Pasterski:2016qvg} and can be written as follows:
\begin{align}
\tilde{\mathcal{A}}_3\propto g\left(\prod_{i=1}^3 \int_{H_3}[d\hat{p}_i(y_i,w_i)] G_{\Delta_i}\left(\hat{p_i}(y_i,w_i) ; z_i\right) \right) \delta^{(4)}\left(-2(1+\varepsilon) \hat{p}_1+\hat{p}_2+\hat{p}_3\right).
\end{align}
Leading order about $\varepsilon$ (mass conservation limit) of this celestial amplitude  can be written as follows \cite{Pasterski:2016qvg}:
\begin{align}
\tilde{\mathcal{A}}_3\propto \sqrt{\varepsilon}\int_{H_3} [d\hat{p}(y,w)] \prod_{i=1}^3G_{\Delta_i}\left(\hat{p}(y,w) ; z_i\right).
\end{align}
This is the contact Witten diagram on EAdS$_3$, and this imply celestial amplitude can be written as AdS$_3$ correlator. If considering sub$^n$leading part, we can expect additional Witten diagram contribute. However, the essential claim of the above discussion holds. Discussion about relations between more general class of celestial amplitude and EAdS$_3$ correlator can be seen in \cite{Cardona:2017keg}.

%%%%%%%%%%%%%%%%%%%%%%%%%%%%%%%%
\section{$H^+_3$-WZW model and AdS$_3$/CFT$_2$} \label{sec: WZW_Review}
    Here we begin with a review of the $H^+_3$-WZW model and discuss its relation to EAdS$_3$/CFT$_2$ correspondence in order to specify a certain part of the model with the CCFT in the next section. In section \ref{subsec: 3.2}, a good omen of selecting the model is provided with the aid of a concrete example of AdS$_3$/CFT$_2$ correspondence.
    
    \subsection{$H^{+}_3$-WZW model}
        \paragraph{General description of WZW models.} 
            The following discussion is just to summarize the general features so those who understand may skip this section. See e.g. \cite{Eberhardt:2019, Walton:1999xc, DiFrancesco:1997nk} for more basics.
            \par The \textit{Wess-Zumino-Witten} (\textit{WZW}) \textit{model} for a generic, possibly non-compact, Lie group $G$ is a conformal nonlinear sigma model $g:S^2\cong\hat{\mathbb{C}}\ni z\mapsto g\qty(z)\in G$ such that the classical action is
            \begin{align}
                \begin{aligned}
                    S_{\text{WZW}}\qty[g]=&-\frac{ik}{2\pi}\int_{S^2}d^2z\tr\qty[\partial g^{-1}\bar{\partial}g] \\
                    & \qquad\qquad\qquad-\frac{ik}{24\pi}\int_{B^3}d^3y\ \epsilon^{ijk}\tr\qty[\qty(\tilde{g}^{-1}\partial_i\tilde{g})\qty(\tilde{g}^{-1}\partial_j\tilde{g})\qty(\tilde{g}^{-1}\partial_k\tilde{g})],
                \end{aligned}
            \end{align}
            where $k$ is the \textit{level} of the theory. The second term, so-called \textit{Wess-Zumino term}, is necessary for the theory to be conformal even after quantization \cite{Witten:1983ar}. It includes the extension $\tilde{g}:B^3\ni y\mapsto \tilde{g}\qty(y)\in G$ of $g$, so the integration over $B^3$ makes sense but the existence of $\tilde{g}$ is essentially non-trivial. Indeed, the obstruction is topological and controlled by the second homotopy class of the target space $G$, favorably vanishing in the case of $H^+_3\cong SL\qty(2,\mathbb{C})/SU\qty(2)$ (see appendix \ref{sec:append_A}). 
            
            The holomorphic and antiholomorphic currents are
            \begin{align}
                J\qty(z):=kg\qty(z)\partial g^{-1}\qty(z),\ \bar{J}\qty(z):=kg\qty(z)\bar{\partial}g^{-1}\qty(z), \label{WZW_current}
            \end{align}
            and they have expansions $J\qty(z)=\sum_a J^a\qty(z)T^a,\ \bar{J}\qty(z)=\sum_a \bar{J}^a\qty(z)T^a$ with respect to the basis $\qty(T^a)_{a}$ of the Lie algebra $\mathfrak{g}=\operatorname{Lie}G$ satisfying $\qty[T^a,T^b]=if^{ab}_cT^c$. The OPE among $J^a\qty(z)$'s are found by the \textit{conformal Ward identity} as
            \begin{align}
                J^a\qty(z)J^b\qty(w)\sim\frac{k\delta^{ab}}{\qty(z-w)^2}+\frac{if^{ab}_c}{z-w}J^c\qty(w).
            \end{align}
            Here, the symbol ``$\sim$'' implies that only the singular part of the Laurent expansion is extracted. 
            $J^a$ have mode expansions $J^a\qty(z)=\sum_{n\in\mathbb{Z}}J^a_nz^{-n-1}$
            and the algebra consisting of the coefficients $\qty(J^a_n)_{n\in\mathbb{Z}}$ reads
            \begin{gather}
                \qty[J^a_m,J^b_n]=km\delta^{ab}\delta_{m+n,0}+if^{ab}_cJ^c_{m+n}, \\
                \qty(J^a_n)^{\dagger}=-J^a_{-n}, \label{hermitian_conjugate}
            \end{gather}
            for compact Lie groups. The Hermitian conjugate rule \eqref{hermitian_conjugate}, however, is not necessarily the case for non-compact groups.

        \paragraph{$H^+_3$ model.}
            \par The following discussion is based on \cite{Teschner:1997ft, Teschner:1997fv, Teschner:1999ug,Ribault:2014hia} with most of the conventions borrowed from \cite{Ribault:2014hia}. 
            \par The \textit{$H^+_3$-WZW model} is a coset theory obtained by taking $G=SL\qty(2,\mathbb{C})$ and gauging with the subgroup $H=SU\qty(2)$ \cite{Gawedzki:1991yu} whose target space is the 3d hypersurface $H^+_3$ identical to the momentum hyperboloid \eqref{on-shell momentum} in a purely mathematical sense. 
            The Hermitian conjugate rule for the currents \eqref{WZW_current} is
            \begin{align}
            \label{eq: H3AlgConj}
                \qty(J^a_n)^{\dagger}=-\bar{J}^a_{-n}.
            \end{align}
            \par The protagonists of the model are \textit{affine} or \textit{Ka\v{c}-Moody primary fields} $\phi^j$ each corresponding to the irreducible representations\footnote{In this model only the \textit{continuous representations} are paramount, and we do not take up other types of representations since there appears only continuous representation in the semi-classical Hilbert space (the Hilbert space for general $k$ is constructed from semi-classical one accompanied by $J_{-n}$ decsendants). See section \ref{sec: 4.5}. The $SL(2,\mathbb{R})$-WZW model, a Lorentzian counterpart of $H^+_3$-model, in comparison involves both continuous and discrete representations \cite{Maldacena:2000hw}.} of $SL\qty(2,\mathbb{C})$ labeled by the \textit{spin} $j=-\frac{1}{2}+i\rho\ \qty(\rho\in\mathbb{R})$. They have the following OPE with the currents
            \begin{align}
                J^a\qty(z)\phi^j\qty(x,w)\sim\frac{\mathcal{D}^a_j}{z-w}\phi^j\qty(x,w),\quad \bar{J}^a\qty(\bar{z})\phi^j\qty(x,\bar{w})\sim\frac{\bar{\mathcal{D}}^a_j}{z-w}\phi^j\qty(x,\bar{w}), \label{affine_primaries}
            \end{align}
            where $z,w\in S^2$(worldsheet coordinate) and $x\in \hat{\mathbb{C}}\cong S^2$ is a certain basis label whose importance would later become apparent. The ``coefficients"
            \begin{align}
                \mathcal{D}^+_j\:=-x^2\partial_x+2jx,\quad\mathcal{D}^0_j:=-x\partial_x+j,\quad\mathcal{D}^-_j:=-\partial_x,
            \end{align}
            are indeed differential operator realization of $\mathfrak{sl}\qty(2,\mathbb{C})$ basis $T^{\pm},\ T^0$ ($\qty[T^0,T^{\pm}]=\pm T^{\pm},\ \qty[T^+,T^-]=2T^0$) acting on $L^2\qty(\mathbb{C})$. We cannot be too careful for the distinction between the $z$-coordinate $S^2$ and the $x$-coordinate $S^2$ in the subsequent analysis. Under the global $SL\qty(2,\mathbb{C})$ action, the transformation laws for $\phi^j$'s are
            \begin{align}
                \phi^j\qty(\frac{\alpha x+\beta}{\gamma x+\delta},z)=|\gamma x+\delta|^{-4j}\phi^j\qty(x,z). \label{wzw_primary}
            \end{align}
            and via the Sugawara construction 
            \begin{align}
                T\qty(z):=-\frac{1}{k-2}\sum_a:J^aJ^a:\qty(z),
            \end{align}
            they turn into Virasoro primary fields
            \begin{align}
                T\qty(z)\phi^j\qty(x,w)\sim\frac{h_j}{\qty(z-w)^2}\phi^j\qty(x,w)+\frac{1}{z-w}\partial\phi^j\qty(x,w),
            \end{align}
            with conformal weights
            \begin{align} \label{spacetime_weight}
                h_j:=-\frac{1}{k-2}j\qty(j+1)=\frac{1}{k-2}\qty(\rho^2+\frac{1}{4}).
            \end{align}
            The central charge computed from the OPE between the two energy-momentum tensors is
            \begin{align}
                c_{H^+_3}=\frac{3k}{k-2}.
            \end{align}
            \par By taking semi-classical limit $k\rightarrow\infty$, namely the \textit{mini-superspace limit}, the correlation function approaches \cite{Ribault:2014hia}
            \begin{align}
                \lim_{k\rightarrow\infty}\expval{\prod^N_{i=1}\phi^{j_i}\qty(x_i,z_i)}\propto\int_{H^+_3}dg\prod^N_{i=1}\psi^{j_i}\qty(x_i,g),
            \end{align}
            where the Haar measure $\int_{H^+_3} dg$ is the same as the one appeared in \eqref{conformal_primary} and 
            \begin{align}
            \label{eq: H3classicallimit}
                \psi^j\qty(x,g)=\frac{2j+1}{\pi}\qty(\qty(
                    \begin{matrix}
                        x \\
                        1
                    \end{matrix}
                )^{\dagger}g\qty(
                    \begin{matrix}
                        x \\
                        1
                    \end{matrix}
                )
                )^{2j}=\frac{2j+1}{\pi}\qty(\frac{y}{y^2+|x-w|^2})^{1-2i\rho}
            \end{align}
            is a function on $H^+_3$ proportional to the celestial bulk-boundary operator \eqref{bulk_boundary} with $\lambda=-2\rho$ ($y$ and $w$ come from the same parametrization \eqref{on-shell momentum} of $H^+_3$). On the basis of this fact, $x\in S^2$ is not merely a label but a coordinate on the boundary of $H^+_3$. 
            \par With the above setups, the two-/three-point functions for spins $j_i=-\frac{1}{2}+i\rho_i$ are given by
            \begin{align}
                &
                \begin{aligned} \label{wzw_two_point}
                    & \ev*{\phi^{j_1}\qty(x_1,z_1)\phi^{j_2}\qty(x_2,z_2)} \\
                    & \qquad\qquad=\frac{1}{|z_1-z_2|^{2h_{j_2}}}\Bigg(-\frac{\pi^2b}{\qty(2j_1+1)^2}\delta\qty(\rho_1+\rho_2)\delta^{\qty(2)}\qty(x_1-x_2) \\
                    & \qquad\qquad\qquad\qquad\qquad\qquad\qquad\qquad+\frac{\pi}{b}\gamma\qty(-\frac{2j_1+1}{b})\delta\qty(\rho_1-\rho_2)|x_1-x_2|^{4j_2}\Bigg),  
                \end{aligned} 
                \\
                \nonumber \\
                &
                \begin{aligned} \label{wzw_three_point}
                    & \ev*{\phi^{j_1}\qty(x_1,z_1)\phi^{j_2}\qty(x_2,z_2)\phi^{j_3}\qty(x_3,z_3)} \\
                    & \qquad\qquad=C^{H^+_3}_{j_1,j_2,j_3}\cdot\frac{|z_3-z_2|^{h_{j_1}-h_{j_2}-h_{j_3}}|z_3-z_1|^{h_{j_2}-h_{j_3}-h_{j_1}}|z_2-z_1|^{h_{j_3}-h_{j_1}-h_{j_2}}}{|x_3-x_2|^{2\qty(j_1-j_2-j_3)}|x_3-x_1|^{2\qty(j_2-j_3-j_1)}|x_2-x_1|^{2\qty(j_3-j_1-j_2)}} ,
                \end{aligned}
            \end{align}
            with $b^2:=k-2$ and $C^{H^+_3}_{j_1,j_2,j_3}$ some numerical constants given in the next section.

    \subsection{AdS$_3$ string / symmetric orbifold duality} \label{subsec: 3.2}
        In this subsection, we briefly review the AdS${}_3$ string / symmetric orbifold duality, which is a concrete example of AdS${}_3$/CFT${}_2$, to explain our motivation to consider $H^+_3$-WZW model.
        Actually, $H^+_3$-WZW model itself is NOT what we consider as the toy-model of CCFT: we need to consider the boundary theory of bulk WZW model as CCFT. Readers who can accept this idea may skip here and jump to the next section.

        \paragraph{AdS$_3$ string theory.}
        AdS${}_3$ side is string theory on AdS${}_3\times \mathcal{M}$ and its worldsheet description is $SL(2,\mathbb{R})_k$ WZW model $\otimes$ sigma model with target space $\mathcal{M}$.
        The central charge of the worldsheet CFT is 26 without $bc$ ghosts ($c_{bc}=-26$),
        \begin{align}
        \label{eq:WSCentralCharge}
            c_{\mathrm{AdS}_3}=\frac{3k}{k-2}, ~~~
            c_{\mathcal{M}}=26-\frac{3k}{k-2}.
        \end{align}
        $SL(2,\mathbb{R})_k$ WZW model is a Lorentzian version of $H^+_3$ WZW model, as they are related by the Wick rotation in target AdS${}_3$.

        There needs to be introduced spectrally-flowed sectors in the spectrum of the $SL(2,\mathbb{R})_k$ WZW model as revealed in \cite{Maldacena:2000hw}.
        Spectrally-flowed sectors are labelled by integer $w$.
             
        \paragraph{Symmetric orbifold CFT.}
        Dual boundary CFT is symmetric orbifold CFT\footnote{Actually we need to take a grand canonical ensemble i.e. summation over $N=1,\dots$.}\footnote{As done in \cite{Hikida:2023jyc}, marginal deformation must be introduced for $k>3$ in order to get the correct correspondence.},
        \begin{align}  
            \mathrm{Sym}^N \left( 
            \text{Linear Dilaton theory with }Q=\frac{k-3}{\sqrt{k-2}} 
            \otimes\mathcal{M}\right).
        \end{align} 
        $\mathcal{M}$ denotes the same theory mentioned in worldsheet theory.
        The central charge of the seed theory is ($c_\mathcal{M}$ is the same one in \eqref{eq:WSCentralCharge}),
        \begin{align}
            c_{\text{LD}} =&~ 1+6Q^2 = 1+ \frac{6(k-3)^2}{k-2}, \nonumber\\
            c_{\mathcal{M}} =&~ 26-\frac{3k}{k-2}, \nonumber\\
            c_{\text{total}} =&~ c_{\text{LD}}+c_{\mathcal{M}}
            =6k.
        \end{align}

        Symmetric orbifold CFT is defined to the set of seed CFT and positive integer $N$, as
        \begin{align}
            \mathrm{Sym}^N \left(\mathcal{C}\right) = \mathcal{C}^{\otimes N}/{S_N}.
        \end{align}
        In other words, $N$-th symmetric orbifold CFT is the $S_N$ permutation orbifold of the $N$ times tensor product of seed theory. The central charge of $N$-th symmetric orbifold CFT is $N$ times that of the seed theory.
        
        As is often the case with orbifold CFT, corresponding twisted sector is introduced.
        Twisted sectors are labelled by $S_N$'s conjugacy class.
        Each element of $S_N$ can be written as the product of cyclic permutation. Twisted sector corresponding to single cyclic permutation of length $w$ is called $w$-twisted sector.

        \paragraph{Corresponding quantities.}
        
        String scattering amplitudes and correlation function on dual boundary CFT coincide as shown in (5.18) of \cite{Hikida:2023jyc}.
        In this duality, $w$-spectrally flowed excitation in string theory corresponds to $w$-twist operator insertion in dual CFT side.
        
        Additionally, conformal weight ($L_0$ eigenvalue) in the dual boundary CFT corresponds to $J_0^3$ eigenvalue in the worldsheet CFT (see e.g. section 2.1 of \cite{Eberhardt:2021vsx}).

%%%%%%%%%%%%%%%%%%%%%%%%%%%%%%%%
\section{Celestial CFT from $H^{+}_3$-WZW model} \label{sec: CalcCorrelation}
    In this section, we explain how the $H^+_3$-model realizes the CCFT. Section \ref{sec: 4.1} describes the logic for identifying the boundary of the WZW target space $H^+_3$ with CCFT based on the fact that $x$ is a boundary coordinate accompanied by the idea of symmetric orbifold explained in the previous section. 
    Section \ref{sec: 4.2} rewrites the WZW spin $-\frac{1}{2}+i\rho$ to celestial principal series $\Delta=1+i\lambda$ and construct operator set of CCFT using affine primaries \eqref{affine_primaries}. In section \ref{sec: 4.3} and \ref{sec: 4.4}, we compute two-/three-point functions following the construction and show that the three-point function vanishes. The level $k$ is not limited to a specific value throughout this section, but in section \ref{sec: 4.5} we propose $k\rightarrow\infty$ limit as another potential way to remove $z$ dependence and get a Hilbert space of semi-classical theory.
    
    \subsection{Assumption and main claim} \label{sec: 4.1}
        Motivated from the AdS${}_3$ string / symmetric orbifold duality, we assume that there exists a dual boundary CFT that corresponds to $H^+_3$-WZW string theory's non-spectrally-flowed sector. The coordinates of the dual boundary CFT coincides with the label of $x$-basis, as one can observe its classical limit in \eqref{eq: H3classicallimit}.
    
\begin{figure}[h] \label{figoverview}
            \centering
            \includegraphics[scale = 0.6]{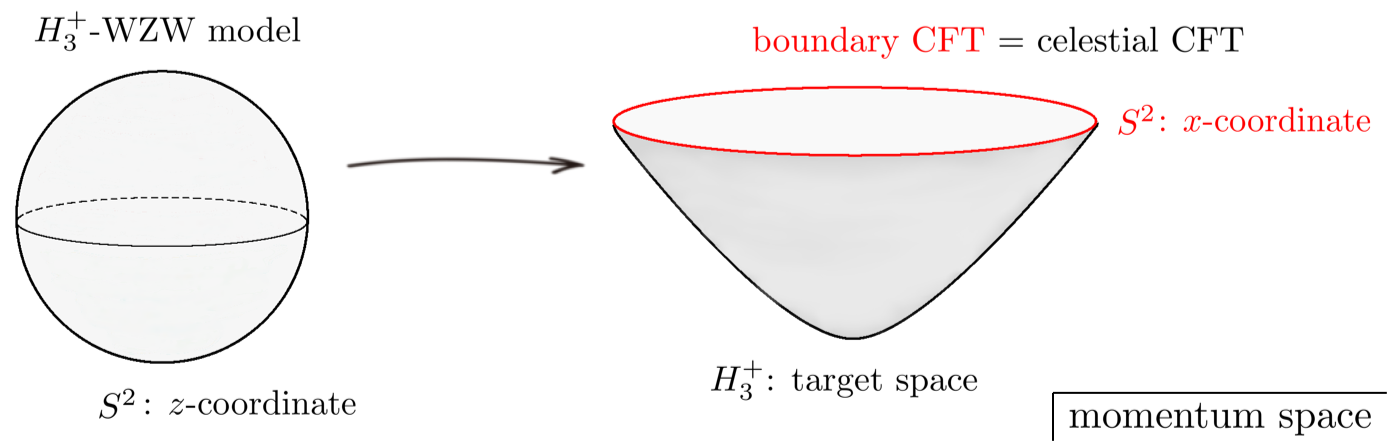}
            \caption{The $H^+_3$-WZW worldsheet and the CCFT as the boundary CFT of the target space $H^+_3$}
            \label{fig:overview}
        \end{figure}

        Considering that massive on-shell particle in Minkowski spacetime consists in $H^+_3$ slice in momentum space, we propose that the CCFT is the dual boundary CFT of $H^+_3$-WZW string theory (Fig. \ref{fig:overview}).
    
        In string theory, $z$ dependence is integrated out. Thus we drop the $z$ dependence (worldsheet coodinate) of the $H_3^+$ correlator to extract the $x$ dependence of string amplitude hereafter. For simplicity, our interest in this section is just sphere contribution for string amplitude, which might correspond to tree-level contribution to the Minkowski scattering amplitude.
    
        Here we should note some general features of this dual CFT, assuming the correspondence.
        The WZW primary fields we consider belong to continuous representation,  which
        correspond to tachyonic excitations in LAdS${}_3$ sense (see section 4.4 of \cite{Maldacena:2000hw}). Thus, our model appears to be non-unitary and fulfills one of the expected characteristics of CCFT.
        We further discuss their semi-classical limit in section \ref{sec: 4.5} and appendix \ref{classical}.
        
        Roughly speaking, current algebra on the worldsheet is translated into boundary Virasoro algebra, at least $J_{0,\pm 1}$ or $L_{0,\pm 1}$ level. Combining it with \eqref{eq: H3AlgConj}, we expect our dual CFT to satisfy
        \begin{align}
            (L_n)^\dagger = - \Bar{L}_{-n}~~~~~(n=0,\pm 1),
        \end{align}
        which is consistent with CCFT's one (see eq. (6.25) of \cite{puhmcelestial}, or \cite{Pasterski:2021fjn}).

    \subsection{Spectrum} \label{sec: 4.2}
        The celestial conformal primary operator $\mathcal{O}_{\Delta=1+i\lambda}\ \qty(\lambda\in\mathbb{R})$ transforms under the global $SL\qty(2,\mathbb{C})$ action as
        \begin{align}
            \qty(g\cdot\mathcal{O}_{\Delta})\qty(x):=\mathcal{O}_{\Delta}\qty(\frac{\alpha x+\beta}{\gamma x+\delta})=|\gamma x+\delta|^{2\Delta}\mathcal{O}_{\Delta}\qty(x),\quad g=\qty(
                \begin{matrix}
                    \alpha & \beta \\
                    \gamma & \delta
                \end{matrix}
            )\in SL\qty(2,\mathbb{C}). \label{ccft_primary}
        \end{align}
        Comparing \eqref{wzw_primary} and \eqref{ccft_primary}, one can infer that $x\qty(\in\mathbb{C})$ coordinate is identical to the celestial sphere coordinate and $\phi^{j=-\frac{1}{2}+i\rho}$ corresponds to $\mathcal{O}_{\Delta=1-2i\rho}$, i.e. there are relations between the principal series and the WZW spin as follows:
        \begin{align} \label{label_correspondence}
            j=-\frac{1}{2}+i\rho\ \longleftrightarrow\ \Delta_i=1-2i\rho=1+i\lambda\ ~~~\qty(\lambda\equiv-2\rho).
        \end{align}
        This is natural and consistent with the fact that conformal weight ($L_0$ eigenvalue) in the dual boundary CFT corresponds to $J^3_0$ eigenvalue in the worldsheet CFT, referred to in section \ref{subsec: 3.2}.
        
        Hereafter, we write the WZW operator $\phi^{\Delta}$ and conformal weight $h_{\Delta_i}$ instead of $\phi^j$ and \eqref{spacetime_weight} on the ground of this fact. Then following the correspondence \eqref{label_correspondence} one can rewrite \eqref{wzw_two_point} and \eqref{wzw_three_point} to
        \begin{align}
            &
            \begin{aligned} \label{two_point_H3}
                &\ev*{\phi^{\Delta_1}\qty(x_1,z_1)\phi^{\Delta_2}\qty(x_2,z_2)} \\
                &\qquad\qquad=\frac{2}{|z_1-z_2|^{2h_{\Delta_1}}}\Bigg(-\frac{\pi^2b}{\qty(\Delta_1-1)^2}\delta\qty(\lambda_1+\lambda_2)\delta^{\qty(2)}\qty(x_1-x_2) \\
                &\qquad\qquad\qquad\qquad\qquad\qquad\qquad\qquad+\frac{\pi}{b}\gamma\qty(\frac{\Delta_1-1}{b^2})\delta\qty(\lambda_1-\lambda_2)\frac{1}{|x_1-x_2|^{2\Delta_1}}\Bigg),  
            \end{aligned} 
        \end{align}
        \begin{align}
            &
            \begin{aligned} \label{three_point_correlator}
                &\ev*{\phi^{\Delta_1}\qty(x_1,z_1)\phi^{\Delta_2}\qty(x_2,z_2)\phi^{\Delta_3}\qty(x_3,z_3)} \\
                &\qquad\qquad=C^{H^+_3}_{\Delta_1,\Delta_2,\Delta_3}\cdot\frac{1}{|z_3-z_2|^{2\qty(h_{\Delta_2}+h_{\Delta_3}-h_{\Delta_1})}|z_3-z_1|^{2\qty(h_{\Delta_3}+h_{\Delta_1}-h_{\Delta_2})}|z_2-z_1|^{2\qty(h_{\Delta_1}+h_{\Delta_2}-h_{\Delta_3})}} \\
                & \qquad\qquad\qquad\qquad\qquad\qquad\times\frac{1}{|x_3-x_2|^{\Delta_2+\Delta_3-\Delta_1}|x_3-x_1|^{\Delta_3+\Delta_1-\Delta_2}|x_2-x_1|^{\Delta_1+\Delta_2-\Delta_3}}, 
            \end{aligned}
        \end{align}
        where $\gamma(x):=\Gamma(x)/\Gamma(1-x)$ (Recall also that $b^2=k-2$).
        These equations are much similar to celestial two- and three-point functions \eqref{nonshadow_shadow} and \eqref{nonshadow_nonshadow} and particularly the delta-function regarding $\lambda_i$ in the R.H.S. of the two-point function suggests that the two-point function with the same $\lambda$ signs follows the power law $|x_1-x_2|^{-2\Delta_2}$(the second term) while the two-point function with different $\lambda$ signs follows the delta-function law $\delta^{\qty(2)}\qty(x_1-x_2)$(the first term). Now the affine primaries $\phi^{j_i}$ are primaries of conformal weights 
        \begin{align}
            h_{\Delta_i} =-\frac{1}{k-2}\qty(-\frac{\Delta_i}{2})\qty(-\frac{\Delta_i}{2}+1)=\frac{1}{4\qty(k-2)}\qty(\lambda^2_i+1) \label{x_weight}
        \end{align}
        w.r.t. $z$-plane (worldsheet), but also are (quasi-)primaries of conformal weights
        \begin{align}
            \frac{\Delta_i}{2}=\frac{1}{2}+i\frac{\lambda_i}{2}
        \end{align}
        w.r.t. $x$-plane (boundary of the target space $H^+_3$), which supports the idea that celestial sphere coordinate is not identical to the $z$-plane but to the $x$-plane.

        We proceed with discussion adopting the shadow transformation \cite{Ribault:2014hia}\footnote{
        This shadow transformation $\widetilde{\phi^{\Delta}}^{\mathrm{Tes}}$ we introduced here slightly differs from the definition in \cite{Ribault:2014hia} by the numerical factor $1/2$ so as for the composition of two consecutive shadow transformations to be identity map. 
        } originally introduced in eq. (40) of Teschner's paper \cite{Teschner:1997ft}. The shadow transformation is related to the celestial shadow defined in \eqref{shadow} by
        \begin{align}
            \widetilde{\phi^{\Delta}}^{\mathrm{Tes}}\qty(x):=&~
            \frac{b^2}{2\pi}\gamma\qty(1+\frac{1-\Delta}{b^2})\int_{\mathbb{C}}d^2x^{\prime}\frac{1}{|x-x^{\prime}|^{2\qty(2-\Delta)}}\phi^{\Delta}\qty(x^{\prime}) \nonumber\\
            =&~ \frac{\Delta-1}{2\pi}\frac{\Gamma\qty(1-i\frac{\lambda}{b^2})}{\Gamma\qty(1+i\frac{\lambda}{b^2})}\int_{\mathbb{C}}d^2x^{\prime}\frac{1}{|x-x^{\prime}|^{2\qty(2-\Delta)}}\phi^{\Delta}\qty(x^{\prime}) \nonumber\\
            =&~ e^{i\alpha(\lambda)}\widetilde{\phi^{\Delta}}\qty(x),
            \label{shadow_transf}
        \end{align}
        where $\Delta = 1+ i\lambda$ and $e^{i\alpha(\lambda)}:=\frac{\Gamma\qty(1-i\frac{\lambda}{b^2})}{\Gamma\qty(1+i\frac{\lambda}{b^2})}$.

        We propose the celestial operator $\mathcal{O}_{\Delta}$ as a linear combination of $H^+_3$-WZW affine primaries 
        \begin{align}
            \mathcal{O}_{\Delta}\qty(x):=&~
            \sqrt{\lambda}\sqrt{\frac{i}{2\pi  b}}\qty(e^{-\frac{i\alpha\qty(\lambda)}{2}}\phi^{\Delta}\qty(x)-e^{+\frac{i\alpha\qty(\lambda)}{2}}\widetilde{\phi^{2-\Delta}}\qty(x))\qquad\qty(\sqrt{\lambda}\text{: principal value})\nonumber \\
            =&~ \sqrt{\lambda}\sqrt{\frac{i}{2\pi  b}}e^{-\frac{i\alpha\qty(\lambda)}{2}}\qty(\phi^{\Delta}\qty(x)-\widetilde{\phi^{2-\Delta}}^{\mathrm{Tes}}\qty(x)) \label{operator_proposal}
        \end{align}
        to exhibit that the CCFT correlators are in agreement with the bulk scattering amplitudes, which is the paramount result of this paper. 
        This idea of linear combination is introduced to realize the vanishing of three-point functions among the same particles discussed in the next subsection.

        We should additionally note that the relation between $\mathcal{O}_{\Delta}$ and $\widetilde{\mathcal{O}_{\Delta}}$ for $\lambda>0$:
        \begin{align}
            \begin{aligned}
            \mathcal{O}_{1+i\lambda} =&~ +i \cdot\widetilde{\mathcal{O}_{1-i\lambda}}, \\
            \mathcal{O}_{1-i\lambda} =&~ -i \cdot\widetilde{\mathcal{O}_{1+i\lambda}}. \label{proportionality}
            \end{aligned}
        \end{align}
    
    \subsection{Two-point function} \label{sec: 4.3}
         \par Based on our definition \eqref{operator_proposal} the two-point functions are\footnote{
            Here the shadow transformation is performed with reference to the formula
            \begin{align*}
                \int d^2w\frac{1}{|z_1-w|^{2\Delta}|w-z_2|^{2\qty(2-\Delta)}}=\frac{4\pi^2}{\lambda^2}\delta^{\qty(2)}\qty(z_1-z_2)\qquad\qty(\Delta=1+i\lambda).
            \end{align*}
        }
        \begin{align} \label{two_point_proposal}
            & \ev*{\mathcal{O}_{\Delta_1}\qty(x_1)\mathcal{O}_{\Delta_2}\qty(x_2)} \nonumber \\
            & \qquad\quad=\frac{i\sqrt{\lambda_1}\sqrt{\lambda_2}}{2\pi b}e^{-\frac{i\alpha\qty(\lambda_1)}{2}}e^{-\frac{i\alpha\qty(\lambda_2)}{2}} \nonumber \\
            & \qquad\qquad\qquad\quad\times\qty(\ev*{\phi^{\Delta_1}\phi^{\Delta_2}}_k-\ev*{\widetilde{\phi^{2-\Delta_1}}^{\mathrm{Tes}}\phi^{\Delta_2}}_k-\ev*{\phi^{\Delta_1}\widetilde{\phi^{2-\Delta_2}}^{\mathrm{Tes}}}_k+\ev*{\widetilde{\phi^{2-\Delta_1}}^{\mathrm{Tes}}\widetilde{\phi^{2-\Delta_2}}^{\mathrm{Tes}}}_k) \nonumber \\
            & \qquad\quad=\frac{2\pi i}{\sqrt{\lambda_1}\sqrt{\lambda_2}}\delta\qty(\lambda_1+\lambda_2)\delta^{\qty(2)}\qty(x_1-x_2)+\frac{\delta\qty(\lambda_1-\lambda_2)}{|x_1-x_2|^{2\Delta_1}}.
        \end{align}
        For $\Delta_1=1+i\lambda_1,~\Delta_2=1+i\lambda_2$ ($\lambda_i>0$), only the second term survive
        \begin{align}
            \ev*{\mathcal{O}_{1+i\lambda_1}\qty(x_1)\mathcal{O}_{1+i\lambda_2}\qty(x_2)}=\frac{\delta\qty(\lambda_1-\lambda_2)}{|x_1-x_2|^{2\Delta_1}}, \label{WZW_nonshadow_nonshadow}
        \end{align}
        whereas for $\Delta_1=1+i\lambda_1,~\Delta_2=1-i\lambda_2$ ($\lambda_i>0$) only the first term is non-vanishing and \eqref{proportionality} leads to
        \begin{align}
            \ev*{\mathcal{O}_{1+i\lambda_1}\qty(x_1)\mathcal{O}_{1-i\lambda_2}\qty(x_2)}&=\frac{2\pi i}{\sqrt{\lambda_1}\sqrt{-\lambda_2}}\delta\qty(\lambda_1-\lambda_2)\delta^{\qty(2)}\qty(x_1-x_2) \nonumber \\
            \Longleftrightarrow~~~~~~\ev*{\mathcal{O}_{1+i\lambda_1}\qty(x_1)\widetilde{\mathcal{O}_{1+i\lambda_2}}\qty(x_2)}
            &=\frac{2\pi i}{\lambda_1}\delta\qty(\lambda_1-\lambda_2)\delta^{\qty(2)}\qty(x_1-x_2). \label{WZW_nonshadow_shadow}
        \end{align}
        \eqref{WZW_nonshadow_nonshadow} and \eqref{WZW_nonshadow_shadow} are the very results  \eqref{nonshadow_nonshadow} and \eqref{nonshadow_shadow}.

    \subsection{Three-point function} \label{sec: 4.4}
        \par For any single massive scalar theory on 4d Minkowski spacetime, any on-shell three-point amplitudes should vanish, leading to the expectation that the corresponding three-point function in CCFT is also zero. In this part, we show that a linear combination \eqref{operator_proposal} of non-shadow $\phi^{\Delta}$ and shadow $\widetilde{\phi^{2-\Delta}}$ satisfies this condition.
    
        \par The starting point is the target space three-point correlators \eqref{three_point_correlator} whose OPE coefficients are given by \cite{Ribault:2014hia}
        \begin{align}
            C^{H^+_3}_{\Delta_1,\Delta_2,\Delta_3}:=\frac{\pi b^{-\frac{1}{b^2}\qty(\Delta_1+\Delta_2+\Delta_3-1)-1}\Upsilon_b^{\prime}\qty(0)\Upsilon_b\qty(\frac{\Delta_1}{b})\Upsilon_b\qty(\frac{\Delta_2}{b})\Upsilon_b\qty(\frac{\Delta_3}{b})}{\Upsilon_b\qty(\frac{\Delta_1+\Delta_2+\Delta_3-2}{2b})\Upsilon_b\qty(\frac{\Delta_1+\Delta_2-\Delta_3}{2b})\Upsilon_b\qty(\frac{\Delta_1-\Delta_2+\Delta_3}{2b})\Upsilon_b\qty(\frac{-\Delta_1+\Delta_2+\Delta_3}{2b})},
        \end{align}
        and $\Upsilon_b$ is so-called the \textit{Upsilon function} (see appendix \ref{sec: append_B}). In what follows, the $x$- and $z$-dependent parts are abbreviated for reasons of space limitation as
        \begin{align}
            & P_{\Delta_1,\Delta_2,\Delta_3}\qty(z_1,z_2,z_3):=|z_3-z_2|^{2\qty(h_{j_2}+h_{j_3}-h_{j_1})}|z_3-z_1|^{2\qty(h_{j_3}+h_{j_1}-h_{j_2})}|z_2-z_1|^{2\qty(h_{j_1}+h_{j_2}-h_{j_3})}, \\
            & Q_{\Delta_1,\Delta_2,\Delta_3}\qty(x_1,x_2,x_3):=|x_3-x_2|^{\Delta_2+\Delta_3-\Delta_1}|x_3-x_1|^{\Delta_3+\Delta_1-\Delta_2}|x_2-x_1|^{\Delta_1+\Delta_2-\Delta_3}.
        \end{align}
        For general values of $k$, three-point correlators with one shadow and two non-shadow operators are given by
        \begin{align}
            \begin{aligned}
                & \ev*{\widetilde{\phi^{2-\Delta_1}}^{\mathrm{Tes}}\qty(x_1)\phi^{\Delta_2}\qty(x_2)\phi^{\Delta_3}\qty(x_3)}\\
                & \quad\quad\quad\quad\quad\quad\quad=\frac{C^{H^+_3}_{2-\Delta_1,\Delta_2,\Delta_3}}{P_{\Delta_1,\Delta_2,\Delta_3}\qty(z_1,z_2,z_3)Q_{\Delta_1,\Delta_2,\Delta_3}\qty(x_1,x_2,x_3)}\frac{\Delta_1-1}{b^2}\gamma\qty(\frac{\Delta_1-1}{b^2}) \\
                \\
                & \quad\quad\quad\quad\quad\quad\quad\quad\quad\quad\quad\quad\quad\times\gamma\qty(\frac{\Delta_1-\Delta_2+\Delta_3}{2})\gamma\qty(\frac{\Delta_1+\Delta_2-\Delta_3}{2})\frac{\Gamma\qty(2-\Delta_1)}{\Gamma\qty(\Delta_1)}. 
            \end{aligned}   
        \end{align}
        We have employed the formula \cite{Simmons-Duffin:2012juh}
        \begin{align}
            & \int d^2w\frac{1}{|z_1-w|^{2a}|z_2-w|^{2b}|z_3-w|^{2c}} \nonumber \\
            \nonumber \\
            & \qquad\qquad\quad=\frac{2\pi\Gamma\qty(1-a)\Gamma\qty(1-b)\Gamma\qty(1-c)}{\Gamma\qty(a)\Gamma\qty(b)\Gamma\qty(c)}\cdot\frac{1}{|z_1-z_2|^{2\qty(1-c)}|z_2-z_3|^{2\qty(1-a)}|z_3-z_1|^{2\qty(1-a)}},
        \end{align}
        and $\gamma\qty(x+1)=-x^2\gamma\qty(x)$ here. It further follows using \eqref{upsilon_formula_3}, \eqref{upsilon_formula_1}, and \eqref{upsilon_formula_2}  that the OPE coefficients satisfy
        \begin{align}
        \label{eq:3ptfuncCondition}
            C^{H^+_3}_{2-\Delta_1,\Delta_2,\Delta_3}=C^{H^+_3}_{\Delta_1,\Delta_2,\Delta_3}\cdot b^2\cdot\frac{\gamma\qty(\Delta_1-1)}{\gamma\qty(\frac{\Delta_1-1}{b^2})}\cdot\frac{1}{\gamma\qty(\frac{\Delta_1-\Delta_2+\Delta_3}{2})\gamma\qty(\frac{\Delta_1+\Delta_2-\Delta_3}{2})},
        \end{align}
        and thus
        \begin{align}
            \ev*{\widetilde{\phi^{2-\Delta_1}}^{\mathrm{Tes}}\qty(x_1)\phi^{\Delta_2}\qty(x_2)\phi^{\Delta_3}\qty(x_3)}=\ev*{\phi^{\Delta_1}\qty(x_1)\phi^{\Delta_2}\qty(x_2)\phi^{\Delta_3}\qty(x_3)}.
        \end{align}
        So if we let celestial operator $\mathcal{O}_\Delta$ be \eqref{operator_proposal}, it finally reproduces the expected identity
        \begin{align}
            \ev*{\mathcal{O}_{\Delta_1}\qty(x_1)\mathcal{O}_{\Delta_2}\qty(x_2)\mathcal{O}_{\Delta_3}\qty(x_3)}=0.
        \end{align}
        It should be strongly noted that $\mathcal{O}_{\Delta}$'s are not zero operators since the two-point functions \eqref{two_point_proposal} are non-vanishing.

    \subsection{Mini-superspace limit} \label{sec: 4.5}
        One can in fact omit the $z$-dependence in both two- and three-point functions
        without considering the WZW model as a string theory,
        by taking the mini-superspace limit $k\rightarrow\infty$, as can be seen from \eqref{x_weight}. In this limit, the Hilbert space of the theory precisely matches with
        \begin{align}
            L^2\qty(H^+_3)\cong\bigoplus_{\rho>0}\mathcal{H}_{-\frac{1}{2}+i\rho},
        \end{align}
        whose basis are the conformal primary wave functions $\Psi_{\Delta}$, and the central charge for the $z$-plane CFT approaches $c_{H^+_3}\rightarrow 3$. This is not inconsistent with the proposal $c=i\infty$ in \cite{Pasterski:2022lsl, Ogawa:2022fhy} since this $c_{H^+_3}\rightarrow 3$ central charge is not for the $x$-plane CCFT but for the worldsheet CFT. Additionally, in the mini-superspace limit, \eqref{shadow_transf} approaches the well-known shadow transformation \eqref{shadow} in celestial holography context
        \begin{align}
            \lim_{k\rightarrow\infty}\widetilde{\phi^{\Delta}}^{\mathrm{Tes}}\qty(x)=\frac{\Delta-1}{2\pi}\int_{\mathbb{C}}d^2x^{\prime}\frac{1}{|x-x^{\prime}|^{2\qty(2-\Delta)}}\phi^{\Delta}\qty(x^{\prime}).
        \end{align}
        Also, in this limit, \eqref{eq:3ptfuncCondition} becomes
        \begin{align}
            \label{eq:3ptfuncConditionMSlimit}\lim_{k\rightarrow\infty}\frac{C^{H^+_3}_{2-\Delta_1,\Delta_2,\Delta_3}}{C^{H^+_3}_{\Delta_1,\Delta_2,\Delta_3}} = -\frac{\Gamma(1-\Delta_1)}{\Gamma(\Delta_1-1)}
            \gamma\qty(\frac{\Delta_1-\Delta_2+\Delta_3}{2})\gamma\qty(\frac{\Delta_1+\Delta_2-\Delta_3}{2}).
        \end{align}

        Physically, the worldsheet shrinks up into line in this limit, and is expected to admit a semi-classical particle picture. 
        As briefly mentioned in section \ref{sec: 4.1}, operators in continuous representation ($\phi^{j}$ with $j = -\frac{1}{2} + i\rho$) correspond to
        tachyonic particle in LAdS${}_3$ sense, and actually they become spacelike geodesics in the $k\to\infty$ limit as considered in \cite{Maldacena:2000hw}. As a consistency check, one can see $\Delta=1+i\lambda$ is conformal weight for states that violate BF bound, namely tachyons.
        As expected, actually we can start from semi-classical setup: we discuss semi-classical AdS${}_3$/CFT${}_2$ in appendix \ref{classical}, where we construct similar operator set of CCFT toy-model without assuming any string theory, if we include tachyons i.e. particles that violate the BF bound.

        In the limit, one can find relation to recent  Liouville theoretic approach to CCFT \cite{Stieberger:2022zyk, Stieberger:2023fju, Taylor:2023bzj, Giribet:2024vnk, Melton:2024gyu}. See the detail in the next discussion part.

%%%%%%%%%%%%%%%%%%%%%%%%%%%%%%%%
\section{Conclusion and discussion}\label{sec : CandD}
In this paper, we propose the boundary theory of $H_3^{+}$-WZW model as a toy-model of CCFT. 
We construct primary operators in CCFT from $H^+_3$-WZW affine primary operator, and checked that the spectrum and two- and three-point functions are consistent with conventional results in CCFT.
Some of future directions of our work are below.

\paragraph{Determination of bulk scalar theory.}
We can calculate bulk four-point scattering amplitude for our model by using celestial holography dictionary. We can determine the mass parameter of exchanging particle for the bulk scalar theory by analysing complex structure of scattering amplitudes of maldelstam variable plane. In addition, we may obtain some hint of bulk interaction.
Mass parameter and level $k$ are expected to be related. 

Moreover, In the context of CCFT, we have to distinguish between incoming and outgoing particles in scattering amplitudes. While this distinction may not pose significant issues when considering massive scattering, it becomes problematic when dealing with massless particles, as pointed out in \cite{Furugori:2023hgv}.
These results are our ongoing work.

\paragraph{Central charge of celestial CFT.}
It has some difficulties in identifying the central charge of CCFT due to some facts like that it is codimension-two holography
and/or there are no parameters like AdS radius because we are dealing with flat spacetime.

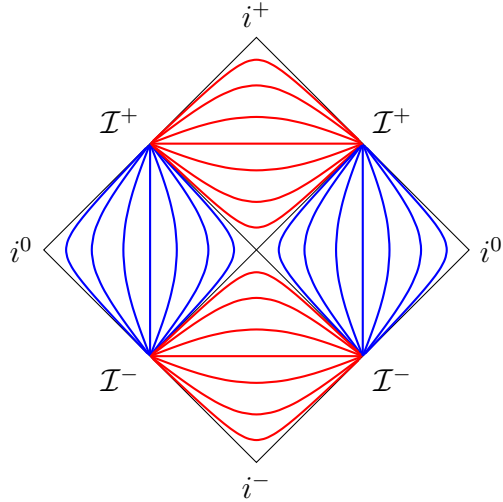
\begin{figure}[htbp]
        \centering

        \begin{tikzpicture}[scale=0.9]
            \coordinate[label = right:$i^0$] (I)   at (pi, 0);
            \coordinate[label = above:$i^+$] (II)  at (0, pi);
            \coordinate[label = left:$i^0$]  (III) at (-pi, 0);
            \coordinate[label = below:$i^-$] (IV)  at (0, -pi);
    
            \draw (I) -- 
            node[midway, above right]    {$\mathcal{I}^+$}
            (II) -- 
            node[midway, above left]    {$\mathcal{I}^+$}
            (III) -- 
            node[midway, below left]    {$\mathcal{I}^-$}
            (IV) -- 
            node[midway, below right]    {$\mathcal{I}^-$} cycle;
    
            \coordinate (M_I_II)   at (pi / 2, pi / 2);
            \coordinate (M_II_III) at (-pi / 2, pi / 2);
            \coordinate (M_III_IV) at (-pi / 2, -pi / 2);
            \coordinate (M_IV_I)   at (pi / 2, -pi / 2);
    
            \draw[very thin] (M_I_II) -- (M_III_IV);
    
            \draw[very thin] (M_II_III) -- (M_IV_I);
    
            \draw[blue, thick] plot[variable = \t, smooth] ({rad(atan(12 * cosh(\t)  / (1 - 36)))},{rad(atan(12 * sinh(\t) / (1 + 36)))});
            \draw[blue, thick] plot[variable = \t, smooth] ({rad(atan(5.4 * cosh(\t)  / (1 - 7.29)))},{rad(atan(5.4 * sinh(\t) / (1 + 7.29)))});
            \draw[blue, thick] plot[variable = \t, smooth] ({rad(atan(3 * cosh(\t)  / (1 - 2.25)))},{rad(atan(3 * sinh(\t) / (1 + 2.25)))});
            \draw[blue, thick] (M_II_III) -- (M_III_IV);
            \draw[blue, thick] plot[variable = \t, smooth] ({-pi + rad(atan((4 / 3) * cosh(\t)  / (1 - (4 / 9))))},{rad(atan((4 / 3) * sinh(\t) / (1 + (4 / 9))))});
            \draw[blue, thick] plot[variable = \t, smooth] ({-pi + rad(atan((2 / 2.7) * cosh(\t)  / (1 - (1 / 7.29))))},{rad(atan((2 / 2.7) * sinh(\t) / (1 + (1 / 7.29))))});
            \draw[blue, thick] plot[variable = \t, smooth] ({-pi + rad(atan((1 / 3) * cosh(\t)  / (1 - (1 / 36))))},{rad(atan((1 / 3) * sinh(\t) / (1 + (1 / 36))))});
    
            \draw[red, thick] plot[variable = \t, smooth] ({rad(atan(12 * sinh(\t)  / (1 + 36)))},{pi + rad(atan(12 * cosh(\t) / (1 - 36)))}) ;
            \draw[red, thick] plot[variable = \t, smooth] ({rad(atan(5.4 * sinh(\t)  / (1 + 7.29)))},{pi + rad(atan(5.4 * cosh(\t) / (1 - 7.29)))});
            \draw[red, thick] plot[variable = \t, smooth] ({rad(atan(3 * sinh(\t)  / (1 + 2.25)))},{pi + rad(atan(3 * cosh(\t) / (1 - 2.25)))});
            \draw[red, thick] (M_III_IV) -- (M_IV_I);
            \draw[red, thick] plot[variable = \t, smooth] ({rad(atan((4 / 3) * sinh(\t)  / (1 + (4 / 9))))},{rad(atan((4 / 3) * cosh(\t) / (1 - (4 / 9))))});
            \draw[red, thick] plot[variable = \t, smooth] ({rad(atan((2 / 2.7) * sinh(\t)  / (1 + (1 / 7.29))))},{rad(atan((2 / 2.7) * cosh(\t) / (1 - (1 / 7.29))))});
            \draw[red, thick] plot[variable = \t, smooth] ({rad(atan((1 / 3) * sinh(\t)  / (1 + (1 / 36))))},{rad(atan((1 / 3) * cosh(\t) / (1 - (1 / 36))))});
    
            \draw[blue, thick] plot[variable = \t, smooth] ({pi + rad(atan(12 * cosh(\t)  / (1 - 36)))},{rad(atan(12 * sinh(\t) / (1 + 36)))});
            \draw[blue, thick] plot[variable = \t, smooth] ({pi + rad(atan(5.4 * cosh(\t)  / (1 - 7.29)))},{rad(atan(5.4 * sinh(\t) / (1 + 7.29)))});
            \draw[blue, thick] plot[variable = \t, smooth] ({pi + rad(atan(3 * cosh(\t)  / (1 - 2.25)))},{rad(atan(3 * sinh(\t) / (1 + 2.25)))});
            \draw[blue, thick] (M_IV_I) -- (M_I_II);
            \draw[blue, thick] plot[variable = \t, smooth] ({rad(atan((4 / 3) * cosh(\t)  / (1 - (4 / 9))))},{rad(atan((4 / 3) * sinh(\t) / (1 + (4 / 9))))});
            \draw[blue, thick] plot[variable = \t, smooth] ({rad(atan((2 / 2.7) * cosh(\t)  / (1 - (1 / 7.29))))},{rad(atan((2 / 2.7) * sinh(\t) / (1 + (1 / 7.29))))});
            \draw[blue, thick] plot[variable = \t, smooth] ({rad(atan((1 / 3) * cosh(\t)  / (1 - (1 / 36))))},{rad(atan((1 / 3) * sinh(\t) / (1 + (1 / 36))))});
    
            \draw[red, thick] plot[variable = \t, smooth] ({rad(atan(12 * sinh(\t)  / (1 + 36)))},{rad(atan(12 * cosh(\t) / (1 - 36)))}) ;
            \draw[red, thick] plot[variable = \t, smooth] ({rad(atan(5.4 * sinh(\t)  / (1 + 7.29)))},{rad(atan(5.4 * cosh(\t) / (1 - 7.29)))});
            \draw[red, thick] plot[variable = \t, smooth] ({rad(atan(3 * sinh(\t)  / (1 + 2.25)))},{rad(atan(3 * cosh(\t) / (1 - 2.25)))});
            \draw[red, thick] (M_I_II) -- (M_II_III);
            \draw[red, thick] plot[variable = \t, smooth] ({rad(atan((4 / 3) * sinh(\t)  / (1 + (4 / 9))))},{- pi + rad(atan((4 / 3) * cosh(\t) / (1 - (4 / 9))))});
            \draw[red, thick] plot[variable = \t, smooth] ({rad(atan((2 / 2.7) * sinh(\t)  / (1 + (1 / 7.29))))},{- pi + rad(atan((2 / 2.7) * cosh(\t) / (1 - (1 / 7.29))))});
            \draw[red, thick] plot[variable = \t, smooth] ({rad(atan((1 / 3) * sinh(\t)  / (1 + (1 / 36))))},{- pi + rad(atan((1 / 3) * cosh(\t) / (1 - (1 / 36))))});
        \end{tikzpicture}

        \caption{Penrose diagram of Minkowski space, the Milne slices. Red lines represent Euclidean AdS$_3$ slices, and blue line is Lorentzian dS$_3$ slices.}
        \label{Milne}
    \end{figure}

There are several proposals regarding the central charge,  
for example, \cite{Pasterski:2022lsl, Ogawa:2022fhy} states $c=i\infty$.
In \cite{Pasterski:2022lsl}, the Virasoro algebra is defined on the Milne slice (Fig.\ref{Milne}) through the application of AdS$_3$/CFT$_2$, allowing for the algebraic definition of the central charge.
In \cite{Ogawa:2022fhy}, extending wedge holography \cite{Akal:2020wfl} to flat spacetimes, the calculation of the partition function of CFT is performed, enabling the determination of the central charge from its cutoff dependence.

On the other hand, \cite{Donnay:2022hkf} show $c=0$ by transformation rules of the stress tensor of CCFT from covariant phase space formalism perspective. \cite{Fotopoulos:2019vac} also show $c=0$ by computing OPE from scattering amplitudes perspective.

It is worth working on determining central charge in our construction. If we naively utilize the AdS$_3$ string / symmetric orbifold duality, central charge in our setup is proportional to $6k$.

\paragraph{Minkowski slicing? Momentum space slicing?}
For one recent approach to celestial holography, there has been a attempt to understand this novel duality as a pile-up of Euclidean AdS${}_3$ and Lorentzian dS${}_3$ \cite{Ball:2019atb}\footnote{
Similar approach to Klein space can be found in \cite{Melton:2023bjw} for example.
}, see Fig.\ref{Milne}.
This idea was firstly introduced by de Boer and Solodukhin \cite{deBoer:2003vf}.

Celestial holography can be regarded as a piling of (A)dS/CFT on such slices. If the bulk is four-dimensional Minkowski spacetime, the slice is either EAdS$_3$ or dS$_3$, and the WZW model has been proposed as a dual CFT to these \cite{Gukov:2004id, Hikida:2022ltr}.

\paragraph{Mini-superspace limit and relation to Liouville theory}

Taking a closer look at AdS$_3$ string / symmetric orbifold duality, the dual boundary CFT of $H^+_3$-WZW we assumed to exist is something related to linear dilaton theory with $c_{\mathrm{LD}}=1+6\frac{(k-3)^2}{k-2}$,
        and this straightly leads to that our model of CCFT is something like linear dilaton.
        This suggests, especially, when we consider the mini-superspace limit $k\rightarrow\infty$ the central charge of our CCFT diverges.

        The authors of \cite{Stieberger:2022zyk, Stieberger:2023fju, Taylor:2023bzj, Giribet:2024vnk, Melton:2024gyu} suggest Liouville theory as CCFT, especially in \cite{Melton:2024gyu}, the Liouville sector of CCFT is proposed to be one with $c_{\mathrm{Liouv}}=1+6(b'+1/b')^2$ in $b'\to0$ limit. In the limit, the potential term in Liouville theory $e^{2b'\phi}$ becomes trivial at least naively and the theory is almost linear dilaton theory.
        Thus the construction in \cite{Melton:2024gyu} is consistent with our approach to CCFT if we take the mini-superspace limit, and parameters are related as $b'\sim1/\sqrt{k}$.

%%%%%%%%%%%%%%%%%%%%%%%%%%%%%%%%

%%%%%%%%%%%%%%%%%%%%%%%%%%%%%%%%
\section*{Acknowledgement}
We are grateful to Ana-Maria Raclariu, Yusuke Taki, Tadashi Takayanagi, and Sotaro Sugishita for instructive discussions. We also appreciate the supervisor of ST, Toshiya Kawai for great advise at the weekly seminar.
The work of NO and TW was supported by JST, the establishment of university fellowships towards the creation of science technology innovation, Grant Number JPMJFS2123.
The work of TT was supported by JST SPRING, Grant Number JPMJSP2110.
The work of NO and TT is partially supported by Grant-in-Aid for Transformative Research Areas (A) “Extreme Universe” No. 21H05187.

\appendix
%%%%%%%%%%%%%%%%%%%%%%%%%%%%%%%%
\section{Basics of $H^+_3\cong SL(2,\mathbb{C})/SU(2)=$ EAdS$_3$} \label{sec:append_A}
    For any invertible matrix $A\in GL\qty(n,\mathbb{C})~\qty(n\in\mathbb{N})$, there is a unique \textit{polar decomposition}
    \begin{align}
        A=U|A|\qquad(\exists! U:\text{ unitary},\ |A|:=\sqrt{A^{\dagger}A}:\text{ the \textit{absolute value} of }A).
    \end{align}
    If one specifies $A$ from $SL\qty(2,\mathbb{C})$, then $\det A=1, |\det U|=1$ and $\det |A|>0$ impliy
    \begin{align}
        \det A=\det U=\det |A|=1,
    \end{align}
    leading to the decomposition as a topological space
    \begin{align}
        SL\qty(2,\mathbb{C})\cong SU\qty(2)\times\qty{\text{positive Hermitian matrices with unit determinant}}.
    \end{align}
    Note that this is not a direct product as a group since the latter space is not closed under the matrix product. It is parameterized by four real numbers $a,b,c,d\in\mathbb{R}$ as
    \begin{align}
        \qty(
            \begin{matrix}
                a+d & b+ic \\
                b-ic & a-d
            \end{matrix}
        ), \label{parametrization}
    \end{align}
    with the constraint
    \begin{align}
        -a^2+b^2+c^2+d^2=-1, \label{eq:hypersurface}
    \end{align}
    arising from the determinant condition. Positiveness also forces two eigenvalues $\alpha_1,~\alpha_2$ of \eqref{parametrization} to be positive, limiting the range of $a$ by
    \begin{align}
        0<\alpha_1+\alpha_2=\tr\qty(
            \begin{matrix}
                a+d & b+ic \\
                b-ic & a-d
            \end{matrix}
        )=2a\iff a>0. \label{positiveness}
    \end{align}
    A widely accepted notation is
    \begin{align}
        H^+_3:=SL\qty(2,\mathbb{C})/SU\qty(2),
    \end{align}
    in accordance with the defining equation \eqref{eq:hypersurface} and \eqref{positiveness} of the \textit{3d hyperbolic surface}, or \textit{Euclidean AdS$_3$ space} (EAdS$_3$). Again from another perspective, the hypersurface has no canonical group structure because the Lie algebra $\mathfrak{sl}\qty(2,\mathbb{C})$ is semisimple and therefore $SL\qty(2,\mathbb{C})$ never has continuous normal subgroups. Since one can easily see that $H^+_3$ is homeomorphic to contractible space $\mathbb{R}^3$, it follows
    \begin{align}
        \pi_2(H^+_3)\cong 0,\ \pi_3(H^+_3)\cong 0,
    \end{align}
    which guarantee WZ-term of the $H^+_3$-WZW model is well-defined, though the level is not necessarily quantized ($k\in\mathbb{Z}$) in contrast to the case of compact Lie groups.

%%%%%%%%%%%%%%%%%%%%%%%%
\section{Mini-superspace limit: semi-classical AdS$_3$/CFT$_2$}\label{classical}

In this appendix, we consider the analytic continuation to $\Delta = 1 + i \lambda$ (breaking the BF bound!) in the traditional AdS$_3$/CFT$_2$ setup and show that a construction of CCFT toy-model similar to that demonstrated in the body of this paper is valid, without considering any string theory. Actually, the fact that celestial operators violate the BF bound in the sense of AdS$_3$/CFT$_2$ is already suggested in \cite{Ogawa:2022fhy} by numerical analysis.

From equation (20) and (25) of \cite{Freedman:1998tz}, a contribution from most simple vertex $\mathcal{L}_I(\Delta_1,\Delta_2,\Delta_3)=\phi_{\Delta_1}^{\text{s.c.}}\phi_{\Delta_2}^{\text{s.c.}}\phi_{\Delta_3}^{\text{s.c.}}$ can be calculated as
\begin{align}
    \left\langle\phi_{\Delta_1}^{\text{s.c.}}(x_1)\phi_{\Delta_2}^{\text{s.c.}}(x_2)\phi_{\Delta_3}^{\text{s.c.}}(x_3)\right\rangle_{\text{s.c.}} = \frac{C^{\text{s.c.}}_{\Delta_1\Delta_2\Delta_3}}{|x_1-x_2|^{\Delta_1+\Delta_2-\Delta_3}|x_2-x_3|^{\Delta_2+\Delta_3-\Delta_1}|x_3-x_1|^{\Delta_3+\Delta_1-\Delta_2}}, \nonumber\\
    C^{\text{s.c.}}_{\Delta_1\Delta_2\Delta_3} = -
    \frac{
    \Gamma\left[\frac{\Delta_1+\Delta_2-\Delta_3}{2}\right]
    \Gamma\left[\frac{\Delta_2+\Delta_3-\Delta_1}{2}\right]
    \Gamma\left[\frac{\Delta_3+\Delta_1-\Delta_2}{2}\right]
    }
    {2\pi^2\Gamma\left[\Delta_1-1\right]\Gamma\left[\Delta_2-1\right]\Gamma\left[\Delta_3-1\right]}
    \Gamma\left[\frac{\Delta_1+\Delta_2+\Delta_3-2}{2}\right],
\end{align}
where ``s.c.'' stands for semi-classical.
Traditionally we impose $\Delta = 1 + \sqrt{1+(m/R_{\text{AdS}})^2}\geq1$, in other words $m$ must satisfy the BF bound $(m/R_{\text{AdS}})^2\geq-1$.

One can immediately check that the three-point coefficient $C^{\text{s.c.}}_{\Delta_1\Delta_2\Delta_3}$ satisfies
\begin{align}
     \frac{C^{\text{s.c.}}_{\Delta_1\Delta_2\Delta_3}}{C^{\text{s.c.}}_{(2-\Delta_1)\Delta_2\Delta_3}}= \frac{\Gamma(1-\Delta_1)}{\Gamma(\Delta_1-1)}
            \gamma\qty(\frac{\Delta_1-\Delta_2+\Delta_3}{2})\gamma\qty(\frac{\Delta_1+\Delta_2-\Delta_3}{2}),
\end{align}
whose right-hand side is $-1$ times that of \eqref{eq:3ptfuncConditionMSlimit}.  
If we set the celestial operator $\mathcal{O}^{\text{s.c.}}_\Delta$ as 
\begin{align}
\label{eq:DefCelOPinSC}
    \mathcal{O}^{\text{s.c.}}_\Delta(x) \propto \left( \phi_{\Delta}^{\text{s.c.}}(x) + \widetilde{\phi_{2-\Delta}^{\text{s.c.}}}(x) \right),
\end{align}
$\mathcal{O}^{\text{s.c.}}_\Delta$ enjoys the vanising three-point function, namely
\begin{align}
    \ev*{\mathcal{O}^{\text{s.c.}}_{\Delta_1}\qty(x_1)\mathcal{O}^{\text{s.c.}}_{\Delta_2}\qty(x_2)\mathcal{O}^{\text{s.c.}}_{\Delta_3}\qty(x_3)}_{\text{s.c.}}=0.
\end{align}
After the analytic continuation $\Delta = 1 + i \lambda$ and if we properly take the normalization constant in \eqref{eq:DefCelOPinSC}, $\mathcal{O}^{\text{s.c.}}_\Delta$s satisfy celestial two-point functions \eqref{nonshadow_nonshadow} and \eqref{nonshadow_shadow}. All other arguments for massive scalar CCFT mentioned in the body of this paper are also satisfied.

Our claim in this appendix is, the CFT side of semi-classical AdS$_3$/CFT$_2$ with spectrum $\phi^{\text{s.c.}}_{1+i\lambda}$ for all $\lambda\in\mathbb{R}$ (infinite number of tachyons) and interaction term $\int_{\lambda_1,\lambda_2,\lambda_3\in\mathbb{R}}\mathcal{L}_I(1+i\lambda_1,1+i\lambda_2,1+i\lambda_3)$
is an alternative toy-model of CCFT.

%%%%%%%%%%%%%%%%%%%%%%%%
\section{Some useful formulae for Upsilon function}\label{sec: append_B}
    Although there are many references for formulae of these sorts, we mainly refer to appendix A of \cite{Nakayama:2004vk} for its conciseness.
    The \textit{Upsilon function} $\Upsilon_b(z)$ is defined as
    \begin{align}
        \Upsilon_b(z):=\frac{1}{\Gamma_b(z)\Gamma_b(b+b^{-1}-z)}.
        \label{Upsilon_def}
    \end{align}
    using the special function
    \begin{align}
        \Gamma_b\qty(z):=\frac{\Gamma_2\qty(z|b,b^{-1})}{\Gamma_2\qty(\frac{1}{2}(b+b^{-1})|b,b^{-1})}
    \end{align}
    Here, $\Gamma_2(z|\alpha,\beta)$ is the \textit{double gamma function}
    \begin{align}
        \Gamma_2(z|\alpha,\beta):=\exp\qty(\left.\frac{\partial}{\partial s}\zeta_2(s,z|\alpha,\beta)\right|_{s=0})\qquad\qty(\Re\alpha,\ \Re\beta>0),
    \end{align}
    and $\zeta_2(s,z|\alpha,\beta)$ is so-called the \textit{Barnes double zeta function}, namely
    \begin{align}
        \zeta_2(s,z|\alpha,\beta):=\sum_{n_1,n_2=0}^\infty \frac{1}{\qty(z+\alpha n_1+\beta n_2)^s}\qquad\qty(\Re s>2,\ \Re z>0),
    \end{align}
    meromorphically continued to all complex $s$ with only singularities simple poles at $s=1,\ 2$. From the definition \eqref{Upsilon_def}, it immediately follows that
    \begin{align} \label{upsilon_formula_3}
        \Upsilon_b\qty(z)=\Upsilon_b\qty(b+b^{-1}-z).
    \end{align}
    Other powerful fomulae are obtained by starting with
    \begin{align}
        \zeta_2(s,z+b|b,b^{-1})&=\sum_{n_1,n_2=0}^\infty \frac{1}{\qty(z+b+b n_1+b^{-1} n_2)^s}\nonumber\\
        &=\sum_{n_1,n_2=0}^\infty \frac{1}{\qty(z+b n_1+b^{-1} n_2)^s}-\sum_{n_2=0}^\infty \frac{1}{\qty(z+b^{-1} n_2)^s}\nonumber\\
        &=\zeta_2(s,z|b,b^{-1})-b^{s}\zeta(s,bz),
    \end{align}
    where $\zeta(s,a)$ is the \textit{Hurwitz zeta function}
    \begin{align}
        \zeta(s,a):=\sum_{n=0}^{\infty}\frac{1}{\qty(n+a)^s}\qquad\qty(\Re s>1,\ \Re a>0),
    \end{align}
    that has a meromorphic continuation to the whole complex plane whose only singularity is simple pole at $s=1$. Then the well-known facts
    \begin{gather}
        \zeta\qty(0,a)=\frac{1}{2}-a, \\
        \left.\frac{\partial}{\partial s} \zeta(s, a)\right|_{s=0}=\log \Gamma(a)-\frac{1}{2} \log (2 \pi),
    \end{gather}
    validate the computation of the transformation law
    \begin{align}
        \Gamma_2(z+b|b,b^{-1})&=\exp\qty(\left.\frac{\partial}{\partial s}\zeta_2(s,z+b|b,b^{-1})\right|_{s=0})\nonumber\\
        &=\exp\qty(\left.\frac{\partial}{\partial s}\zeta_2(s,z|b,b^{-1})\right|_{s=0}-\left.b^s\log{b}\cdot\zeta(s,bz)\right|_{s=0}-\qty( \log{\Gamma(bz)}-\frac{1}{2}\log(2\pi) ))\nonumber\\
        &=\exp\qty(\left.\frac{\partial}{\partial s}\zeta_2(s,z|b,b^{-1})\right|_{s=0}-\qty(\frac{1}{2}-bz)\log{b}- \log{\Gamma(bz)}+\frac{1}{2}\log(2\pi) )\nonumber\\
        &=\frac{\sqrt{2\pi}b^{bz-\frac{1}{2}}}{\Gamma(bz)}\Gamma_2(z|b,b^{-1}),
    \end{align}
    which by replacing $z\rightarrow b^{-1}-z$ entails another equation
    \begin{align}
        \Gamma_2(b^{-1}-z+b|b,b^{-1})=\frac{\sqrt{2\pi}b^{\frac{1}{2}-bz}}{\Gamma(1-bz)}\Gamma_2(b+b^{-1}-(z+b)|b,b^{-1}).
    \end{align}
    Combining these results, we get a translation law
    \begin{align} \label{upsilon_formula_1}
        \Upsilon_b(z+b) & = \frac{1}{\Gamma_b(b+b^{-1}-(z+b)|b,b^{-1})\Gamma_b(z+b|b,b^{-1})}\nonumber\\
        & =\frac{\Gamma(bz)}{\Gamma(1-bz)}b^{1-2bz}\Upsilon_b(z) \nonumber \\
        & = \gamma\qty(bz)b^{1-2bz}\Upsilon_b(z).
    \end{align}
    A similar manner computation yields another translation law
    \begin{align} \label{upsilon_formula_2}
        \Upsilon_b(z+b^{-1}) & = \gamma(b^{-1}z)b^{2b^{-1}z-1}\Upsilon_b(z).
    \end{align}

\bibliographystyle{JHEP}
\bibliography{bib}

\end{document}